\documentclass[aps,prl,superscriptaddress,reprint]{revtex4-1}
\usepackage{graphicx}
\usepackage[utf8]{inputenc}
\usepackage{amsmath}
\usepackage{amssymb}
\usepackage{hyperref}
\usepackage{url}
\usepackage{color}
\usepackage{cases}

\newcommand{\atilde}{\tilde{a}}
\newcommand{\pitilde}{\tilde{\pi}}
\newcommand{\eps}{\varepsilon}
\newcommand\tab[1]{Table~\ref{#1}}
\newcommand\eq[1]{Eq.~\ref{#1}}
\newcommand\fig[1]{Fig.~\ref{#1}}

\newcommand\bigO[1]{\ensuremath{\OCAL(#1)}}

\newcommand\Var{\operatorname{Var}}
\newcommand\lfree{\ensuremath{\ell_{\text{free}}}}
\newcommand\Lfree{\ensuremath{L_{\text{free}}}}
\newcommand{\VEC}[1]{\mathbf{#1}}
\newcommand{\xvec}{\VEC{x}}
\newcommand{\epsvec}{\boldsymbol{\varepsilon}}
\newcommand{\Umean}{\overline{U}}
\newcommand{\taumix}{\tau_{\text{mix}}}
\newcommand\pacc{P_{\text{acc}}}     
\newcommand\ptrans{T}   
\newcommand\proba{p}    
\newcommand{\ACAL}{\mathcal{A}}  
\newcommand{\FCAL}{\mathcal{F}}  
\newcommand{\OCAL}{\mathcal{O}}  
\newcommand{\RCAL}{\mathcal{R}}  
\newcommand{\glb}{\left(}  
\newcommand{\grb}{\right)}  
\newcommand{\glc}{\left[}  
\newcommand{\grc}{\right]}  
\newcommand{\gld}{\left\{}  
\newcommand{\grd}{\right\}}  

\newcommand{\quot}[1]{``#1''}
\newcommand\figmix{\fig{figMixing}c}

\let\ifincludesupplements\iftrue
\let\ifnotbuildingseparatesupp\iftrue

\date{\today}
\begin{document}

\title{Irreversible local Markov chains with rapid convergence 
towards equilibrium} 
\author{Sebastian C. Kapfer}
\email{sebastian.kapfer@fau.de}
\affiliation{Theoretische Physik 1, FAU Erlangen-N\"{u}rnberg, Staudtstr.\ 7, 91058
Erlangen, Germany}

\author{Werner Krauth}
\email{werner.krauth@ens.fr}

\affiliation{Laboratoire de Physique Statistique, D\'{e}partement de physique
de l'ENS, Ecole Normale Sup\'{e}rieure, PSL Research University, Universit\'{e}
Paris Diderot, Sorbonne Paris Cit\'{e}, Sorbonne Universit\'{e}s, UPMC
Univ. Paris 06, CNRS, 75005 Paris, France}

\affiliation{Department of Physics, Graduate School of Science, The University
of Tokyo, 7-3-1 Hongo, Bunkyo, Tokyo, Japan}
\ifnotbuildingseparatesupp
\begin{abstract}
We study the continuous one-dimensional hard-sphere model and present
irreversible local Markov chains that mix on faster time scales than the
reversible heatbath or Metropolis algorithms. The mixing time scales appear to
fall into two distinct universality classes, both faster than for reversible
local Markov chains.  The event-chain algorithm, the infinitesimal limit of one
of these Markov chains, belongs to the class presenting the fastest decay. For
the lattice-gas limit of the hard-sphere model, reversible local Markov chains
correspond to the symmetric simple exclusion process (SEP) with periodic
boundary conditions. The two universality classes for irreversible Markov
chains are realized by the totally asymmetric simple exclusion process (TASEP),
and by a faster variant (lifted TASEP) that we propose here.  We discuss how
our irreversible hard-sphere Markov chains generalize to arbitrary repulsive
pair interactions and carry over to higher dimensions through the concept
of lifted Markov chains and the recently introduced factorized Metropolis
acceptance rule.
\end{abstract}
\maketitle 
\fi

The hard-sphere model plays a central role in statistical mechanics.
In three spatial dimensions (3D), the classical hard-sphere crystal melts
in a first-order phase transition \cite{KirkwoodMonroe1941,*HooverRee1968},
whereas 2D hard spheres undergo a sequence of two phase transitions that have
been characterized only recently \cite{Alder1962, Bernard2011}.  Hard spheres
have established paradigms for order-from-disorder phenomena driven by the
depletion interaction \cite{Asakura1954,SMAC}, and for 2D
melting with its dissociation of orientational and positional order
\cite{HalperinNelson1978,*Young1979VectorCoulomb}.  The dynamics of the
hard-sphere model has also been the focus of great attention, from the first
algorithmic implementation of Newtonian mechanics through event-driven
molecular dynamics \cite{Alder1957} and the discovery of algebraically
decaying velocity autocorrelations \cite{AlderWainwrightAutocorrelation1970}
to insights into the glass transition \cite{Zamponi2010} and granular materials
\cite{TorquatoReview2010}, and from the first definition of Markov-chain Monte
Carlo dynamics \cite{Metropolis1953} to rigorous convergence rates towards
equilibrium in some special cases \cite{Kannanrapidmixing2003}.

\begin{figure}[b]
\includegraphics[width=\linewidth]{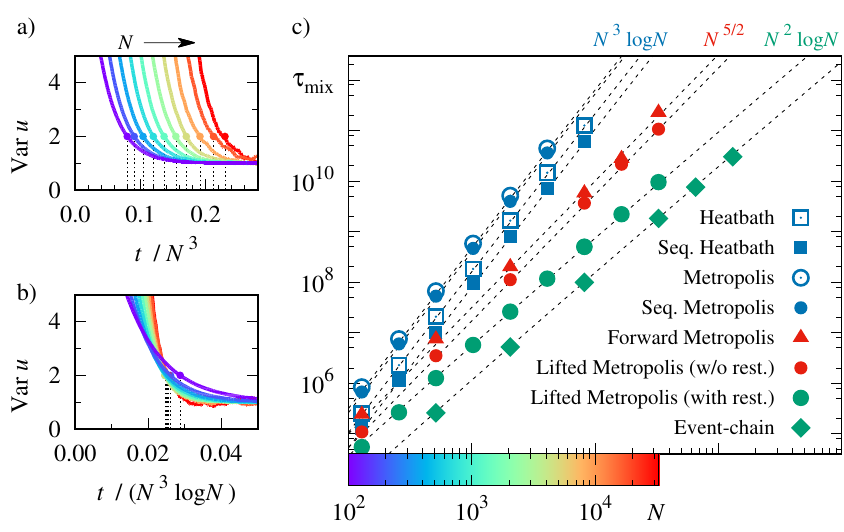}
\caption{Mixing of local 1D Markov chains. a) Relaxation of $\Var u_i$ from
the compact initial state under heatbath dynamics ($x$ axis rescaled by $N^3$,
$y$ axis by equilibrium value).  b) Rescaling of $x$ axis with an additional
logarithm illustrates $\bigO{N^3\log N}$ time scale.  c) Mixing times for the
Markov chains discussed in this work.  The step $\eps$ is uniformly sampled
from $(0;2.5\lfree)$. For the event-chain algorithm, $\taumix$ is measured in
lifting moves.
}
\label{figMixing}
\end{figure}
In 1D, the thermodynamics and the static correlation functions of finite
hard-sphere systems can be computed exactly \cite{Tonks1936, SMAC}. Newtonian
dynamics is pathological for equal sphere masses, because colliding spheres
simply exchange their velocities without mixing them. Stochastic dynamics,
however, may converge to equilibrium.  For example, reversible heatbath
dynamics mixes (that is, converges towards equilibrium from an arbitrary
starting configuration) in at most  $\sim N^3 \log N$ individual steps for $N$
spheres \cite{Abbrev:RandallWinklerCircle2005}.

In the present work, we study irreversible local Markov chains for 1D
hard-sphere systems that violate the detailed-balance condition yet still
converge towards equilibrium.  We show by numerical simulation that these
irreversible Markov chains typically mix faster than reversible ones, and
that they fall into two universality classes (see the Supplemental Material 1
\cite{SUPP} for background on balance conditions and the Supplemental Material
2 for details on mixing and correlation times).  The first one mixes in
\bigO{N^{5/2}} steps, and is related to the totally asymmetric simple exclusion
process (TASEP, see \cite{GwaSpohnPRL1992, ChouTASEP2011, BaikLiu2016}).
The other mixes in $\bigO{N^2\log N}$ and comprises the event-chain algorithm
(ECMC) \cite{Bernard2009} and a modified lifted TASEP that we propose in
this work.  We refer to it as the lifted TASEP class.  The framework for our
approach to irreversible Markov chains is provided by the lifting concept
\cite{Abbrev:Chen1999, Diaconis2000} together with a factorized Metropolis
acceptance rule \cite{Michel2014JCP}, which considerably extend the range of
applications for irreversible Markov chains (see Ref.~\cite{KuKColloquium}
for a review). Some evidence for reduced mixing time scales has already been
obtained \cite{Nishikawa2015}.

In order for a reversible or irreversible Markov chain to converge to the
thermodynamic equilibrium given by $\pi$, the total probability flow $ \FCAL_a$
into a configuration $a$ must satisfy the global balance condition:
\begin{equation}
\FCAL_a
\equiv 
 \sum_b \pi(b) \ptrans(b \to a) = \pi(a),
\label{e:global_balance}
\end{equation}
where $\ptrans(b \to a)$ is the algorithmic transition probability from
$b$ to $a$.  In the following,  we distinguish between \quot{accepted} flow
$\ACAL(b \to a) = \pi(b) \ptrans(b \to a)$ from configurations $b \neq a $ to
$a$  and \quot{rejected} flow $\RCAL=\pi(a)\ptrans(a\to a)$ which results from
an attempted move from $a$ that was not accepted.  The global balance condition
enforces stationarity of $\pi$ under multiplication with
the transfer matrix  $T$ (see the Supplemental Material 1 for
definitions). The special condition realized in reversible Markov chains is
the detailed balance $\pi(b) \ptrans(b \to a) = \pi(a) \ptrans(a \to b)$ which,
in terms of the probability flows is simply $\ACAL(a \to b) = \ACAL(b \to a)$,
and which implies \eq{e:global_balance}.

For concreteness, we restrict ourselves to $N$ hard spheres of diameter $d $
on a circle of length $L$, so that the free space is $\Lfree = L - N d$, and
the mean gap between spheres $\lfree = \Lfree / N$. All valid configurations
$a$ have the same statistical weight $\pi(a) = 1$. They consist in ordered
particle positions $a = (\dots x_{i-1}, x_i, x_{i+1},\dots) $ with gap
variables $\delta_i = x_i - x_{i-1} - d \geq 0$ and appropriate periodic
boundary conditions.  The partition function $Z \sim (\Lfree)^N$ is analytic
for all densities in the thermodynamic limit, and no phase transition takes
place \cite{Tonks1936, SMAC}. The model is isomorphic to $N$ point particles
on a circle of length $\Lfree$ with the same gap variables and an interaction
$V(\delta_i < 0) = \infty $ and $V(\delta_i \geq 0) = 0$ implementing both the
non-overlap and the ordering constraint.  Because of this mapping onto a gas of
free particles with mean gap $\lfree$, and because the step size distribution
of our Markov chains scale with $\lfree$, their dynamics and in particular
their mixing times do not depend on density (see the Supplemental Material 3
for details). Nevertheless,  the spatial correlation
length of the hard-sphere model diverges in the close-packing limit.

We first consider the \emph{reversible heatbath} algorithm, which moves
at each time step $t = 0, 1, \dots $ a random sphere $i$ to a random
position between spheres $i-1$ and $i+1$ ($x_i$ is uniformly sampled in
$(x_{i-1}+d, x_{i+1}-d)$).  When studying the mixing dynamics, we ignore
trivial uniform rotations of the configuration (which only mix in $\sim
N^4$ steps \cite{Abbrev:RandallWinklerCircle2005}).  We thus restrict our
attention to quantities that can be expressed in terms of the $\delta_i$ and
focus on the slow, large-scale density fluctuations.  The reversible heatbath
algorithm is known to mix in at most $\sim N^3\log N$ steps.  We assume that
the slowest time scale is exposed by tracking the distribution $\pi(u_i)$ of
any half-system distance
    $u_i = \delta_i + \delta_{i+1} + \dotsc + \delta_{i+N/2}$
from a compact initial configuration at $t = 0$, where the variance of $u_i$
equals $\Var u_i = \Lfree^2/4$, towards equilibrium at $t \sim \taumix$, where
$\Var u_i = \Lfree^2 / (4N+4)$ (see the Supplemental Material 2 for
details).  Our simulations indeed show, in agreement with the rigorous bounds
\cite{Abbrev:RandallWinklerCircle2005}, that $\sim N^3$ steps are insufficient
for mixing (see \fig{figMixing}a), while $\sim N^3 \log N$ steps suffice
(see \fig{figMixing}b).  The diverging slope of $\Var u_i$ at $\taumix$ signals
the cutoff phenomenon \cite{AldousDiaconis1986}.  Our method also recovers the
correct mixing time for the related discrete symmetric SEP model (see below).
For this model, the leading term of $\taumix$ is known rigorously
\cite{Lacoin_2017_SSEP} and scales as $N^3\log N$ (notwithstanding the absence
of the logarithm in the spectral gap).  Our numerical method thus reliably
detects mixing times, including logarithmic corrections and prefactors.

Plots analogous to \fig{figMixing}a,b obtain the mixing time scales for all
Markov chains studied in the present work (see \figmix).  For the heatbath
algorithm, a simple scaling argument for the discrete lowest-$k$ Fourier mode
$\Umean =  (u_1 + u_2 + \dots + u_{N/2}) / N ^ {3/2}$ yields the \bigO{N^3}
behavior in equilibrium: In the limit $N \to \infty$, the standard deviation of
$\pi(\Umean)$ is \bigO{1}, and one heatbath step changes $\Umean$ by \bigO{1 /
N ^ {3/2}}. A random walk in $t$ yields:
\begin{equation}
\quad \frac{1}{N^{3/2}} \sqrt{t} \sim 1 \implies 
\taumix \gtrsim N^3.
\label{e:scaling_reversible}
\end{equation}

\begin{figure}
\includegraphics[width=\linewidth]{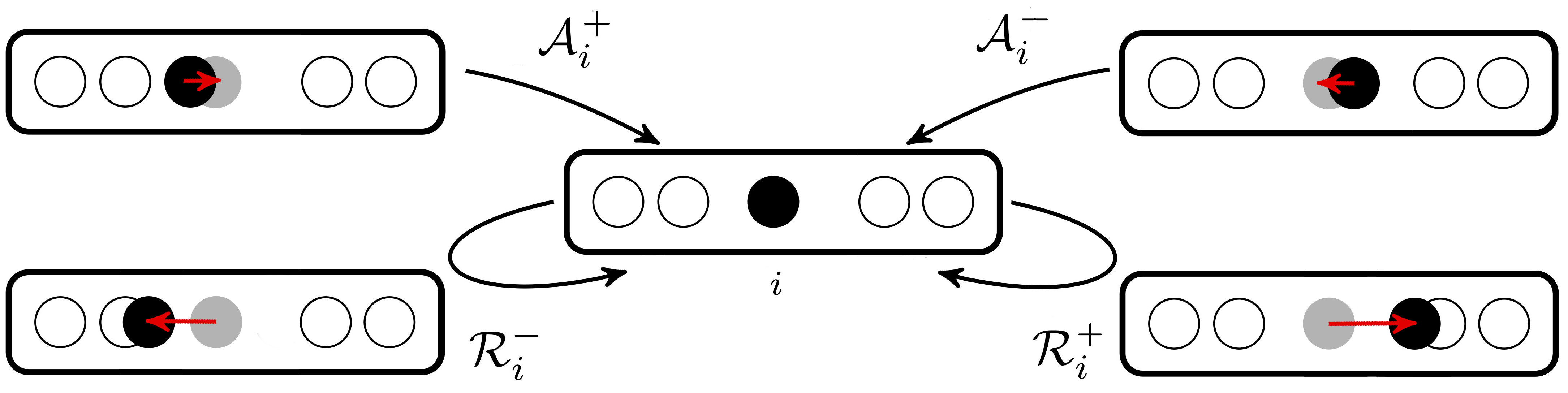}
\caption{
Metropolis flow into configuration $a$ by moves of sphere $i$.  For given
$\eps$, $\glc \ACAL_i^\pm(\eps), \RCAL_i^\mp(\eps) \grc \in \gld [1,0],
[0,1] \grd$ as a move by $\eps$ is either accepted or rejected.  Flows here
are integrated over the step distribution $\proba(\eps)$.  The relation
$\glc \ACAL_i^+(\eps), \RCAL_{i-1}^+(\eps)\grc \in \gld [1, 0], [0, 1] \grd$
justifies the forward Metropolis algorithm.
}
\label{f:Metropolis_global}
\end{figure}
 
The \emph{reversible Metropolis} algorithm mixes on the same \bigO{N^3 \log N}
scale as the reversible heatbath algorithm (see \figmix). At each step, it
considers a move from $a$ towards a configuration $\atilde$ with $\tilde x_i
= x_i + \sigma \eps$, where $\sigma = \pm 1$ samples the forward or backward
direction and $\eps > 0$ samples the step from some distribution $\proba(\eps)$
(we use a uniform distribution on the interval $[0; 2.5\lfree]$).  If $\atilde$
is invalid because of an overlap or an inversion of $x_i$ with $x_{i-1}$ or
$x_{i+1}$, the move $a \to a$ results. The reversible Metropolis algorithm
satisfies detailed balance between $a$ and any $\atilde$ simply because the
moves $\atilde \to a$ and $a \to \atilde$ are equally likely.

The probability flow $\FCAL_a$ into $a$ has four components for each sphere $i$
(see \fig{f:Metropolis_global}), namely accepted forward flow $\ACAL_i^+ =
\int\!\mathrm d\eps\, \proba(\eps) \ACAL_i^+(\eps)$, corresponding to $\sigma
= + 1$ and analogously accepted backward flow $\ACAL_i^-$ for $\sigma = -1$.
Rejected forward and backward flows $\RCAL_i^+$ and $\RCAL_i^-$ from $a$
towards invalid configurations $\atilde$ also contribute to the flow into
$a$. For given $\eps$, these flows,
\begin{equation}
\begin{aligned}
 \ACAL_{i}^+(\eps) &= \Theta( \delta_{i} - \eps ) &
 \RCAL_{i}^+(\eps) &= \Theta( \eps - \delta_{i+1} ) \\
 \ACAL_{i}^-(\eps) &= \Theta( \delta_{i+1} - \eps )\quad &
 \RCAL_{i}^-(\eps) &= \Theta( \eps - \delta_{i} ),
\end{aligned}
\label{e:MetroFlows}
\end{equation}
(with $\Theta$ the Heaviside step function) are either unity or zero.
Moreover, they add up to unity in pairs: $\ACAL_i^+(\eps) + \RCAL_i^-(\eps) =
1$, $\ACAL_i^-(\eps) + \RCAL_i^+(\eps) = 1$, because each move is accepted (or
rejected) under the same condition as its return move.  It follows that for any
distribution of $\eps$, the sum of the four flows equals $2$.

Global balance requires the total flow $\FCAL_a$ into a valid hard-sphere
configuration $a$ to equal $\pi(a) = 1$.  For the reversible Metropolis
algorithm there are $2 N$ equal choices of the $N$ spheres and two directions
$\sigma = \pm 1$, so that
\begin{equation*}
\FCAL_a^{\text{rev}} =     \frac{1}{2N} \sum_i 
    \underbrace{\bigl(\ACAL_{i}^+ + \RCAL_{i}^+ + \ACAL_{i}^- + 
\RCAL_{i}^-\bigr)}_{\text{$=2$, see \eq{e:MetroFlows}}}  = 1.
\label{e:MetroBalance}
\end{equation*}
Global balance is thus established, although it was already implied by the
detailed-balance condition.

Global balance is also satisfied for the \emph{sequential
Metropolis} algorithm, the historically first irreversible Markov chain
\cite{Metropolis1953}, which updates spheres sequentially, say, in ascending
order in $i$. At a given time, only a fixed sphere $i$ is updated, and the flow
into a configuration $a$ during this move arises from the two choices $\sigma =
\pm 1$ for this update of $i$, which each can be either accepted or rejected:
\begin{equation}
\FCAL_a^{\text{seq}} =   \frac{1}{2} \glb \ACAL_{i}^+ + \RCAL_{i}^+ + 
\ACAL_{i}^- + \RCAL_{i}^- \grb = 1.  
\label{e:sequentialMetro}
\end{equation}
Irreducibility and aperiodicity can also be proven for generic distributions
$\proba(\eps)$.  With $\proba(\eps)$ uniform in the interval $[0, 2.5 \lfree]$,
the sequential Metropolis algorithm mixes $\sim 1.2$ times faster than the
reversible Metropolis algorithm, but with the same \bigO{N^3 \log N} scaling.

The relation $\ACAL_i^+(\eps) + \RCAL_i^-(\eps) = 1$  from \eq{e:MetroFlows}
can be expressed  as $\ACAL_i^+(\eps) + \RCAL_{i-1}^+(\eps) = 1$.  This
motivates  the \emph{forward Metropolis} algorithm, which attempts at each
time step a forward move ($\sigma\equiv+1$) sampled from the probability
distribution $\proba(\eps)$, for a randomly sampled sphere $i$.  There are now
$N$ equal choices for the moves, and the incoming flow into a configuration $a$
is given by
\begin{equation}
\FCAL_a^{\text{forw}} = 
    \frac 1N \sum_i
        \underbrace{ \glb \ACAL^+_i + \RCAL^+_{i-1}   \grb }_{=1} = 
    1,
\label{e:forward_Metropolis}        
\end{equation}
which again establishes global balance.  In contrast, the \emph{sequential
forward Metropolis} algorithm violates global balance. In this algorithm,
at a given time step, a fixed sphere $i$ is updated, and the flow
$\FCAL_a^{\text{seq-forw}}$ into a configuration $a$, at this time step,
arises for a given value $\eps$ from a single possible move.  The total flow
is $\FCAL_a^{\text{seq-forw}} = \ACAL^+_i + \RCAL_{i}^+ \neq 1$, so that the
sequential forward algorithm is not correct.

\begin{figure}
\includegraphics[width=\linewidth]{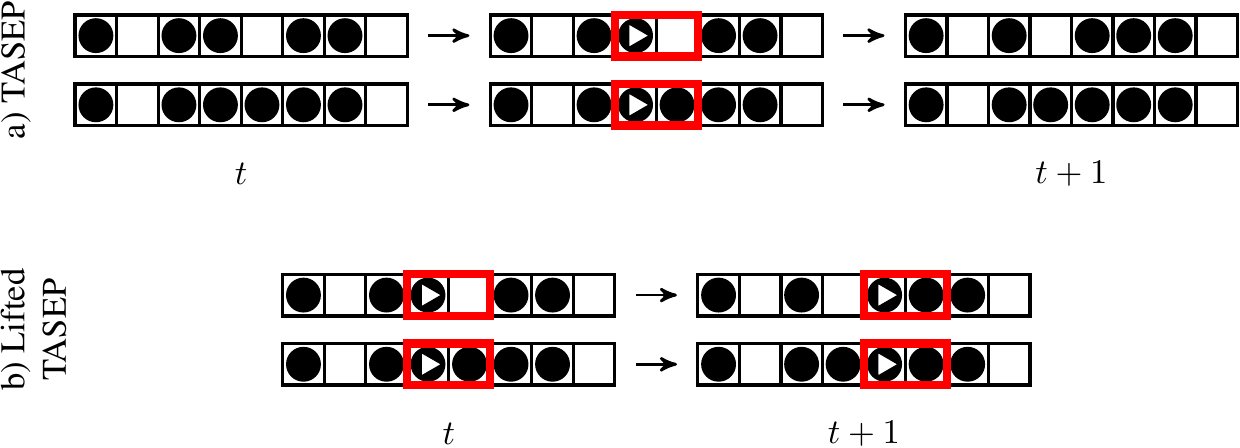}
\caption{Update rules for discrete models. 
a) The TASEP advances a random sphere, if possible, and it mixes in
\bigO{N^{5/2}}.  b) The lifted TASEP, without restarts, is deterministic.
With restarts, it  mixes in \bigO{N^2 \log N }.
}
\label{figAutomatonRules}
\end{figure}

Remarkably, we find that the forward Metropolis algorithm mixes on a time
scale \bigO{N^{5/2}}
(see \figmix{}).  The same $N^{5/2}$ time scale also governs the relaxation
of the totally asymmetric simple exclusion process (TASEP), a lattice
transport model which converges to equilibrium under periodic boundary
conditions \cite{BaikLiu2016}.  An individual TASEP step attempts to move
a randomly-sampled sphere one site to the right (\fig{figAutomatonRules}a).
Indeed, the TASEP agrees with the forward Metropolis algorithm restricted to
integer $x_i$ and $L$ and steps $\eps \equiv 1$.  By tracking the lattice
equivalent of $\Var u_i$, we recover the $\bigO{N^{5/2}}$ mixing for the
TASEP\cite{BaikLiu2016}, while the \emph{symmetric} SEP, itself the lattice
version of the reversible Metropolis algorithm, mixes in $\sim N^3\log N$
\cite{Lacoin_2017_SSEP}.

The relation $ \ACAL^+_i(\eps) + \RCAL^+_{i-1}(\eps) =1$, for any
individual $i$ (see \eq{e:forward_Metropolis}) provides the motivation for
the \emph{lifted Metropolis} algorithm.  Here, moves are attempted in the
forward direction $\sigma\equiv +1$ but
the active sphere $i$ at time step $t+1$ is determined from the
outcome at time step $t$ (see \fig{figAutomatonRules}b for the discretized
example): As long as sphere $i$ can move, it remains active for the next
step. Only when it cannot, a lifting move $i \to i+1$ takes place instead
and $i+1$ becomes the active sphere (the physical configuration $a$ does
not change during this step).  Each configuration is now characterized
by the active particle $i$, in addition to $a$. The incoming flow $
\FCAL_{(a,i)}^{\text{lift}}$ into a lifted configuration $(a,i)$ is either due
to an accepted move of $i$ or a rejected move of $i-1$, so that:
\begin{equation}
\FCAL_{(a,i)}^{\text{lift}} = \ACAL^+_i + \RCAL^+_{i-1} = 1. 
\end{equation}
Global balance again holds.  We find that the lifted Metropolis algorithm,
run as a Markov chain without restarts (see below) mixes in
\bigO{N^{5/2}} steps (see \figmix{}). It thus belongs to the TASEP universality
class.

\begin{figure}[t]
\includegraphics[width=\linewidth]{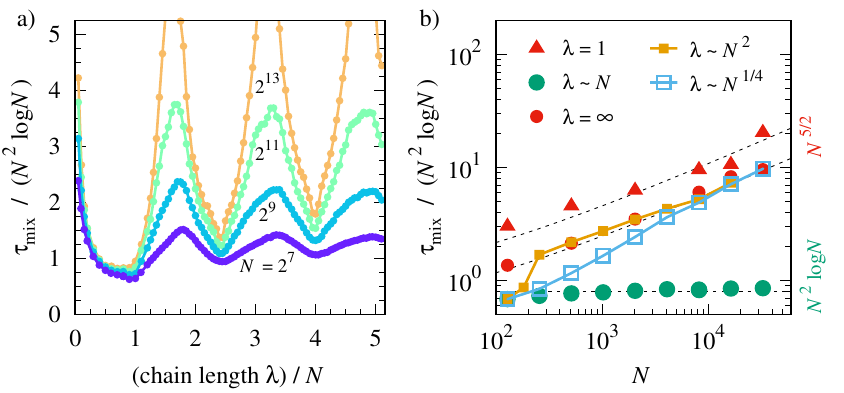}
\caption{Restarts in the lifted Metropolis algorithm. a) Mixing time as
a function of chain length $\lambda \sim N$: The \bigO{N^2 \log N} time 
scale is
preserved but mixing is fastest for $\lambda \simeq 0.9 N$.  b) \bigO{N^{5/2}}
mixing for $\lambda \sim N^{1/4}$ (asymptotically equivalent to the (unlifted)
forward Metropolis algorithm, $\lambda = 1$, red triangles) and for $\lambda
\sim N^2$ (asymptotically equivalent to lifted Metropolis without any restarts,
$\lambda=\infty$, red bullets).
}
\label{figRestartFrequencies}
\end{figure}

Balance conditions relate the stationary probability
distribution $\pi$ at time step $t$ to the distribution at time step
$t+1$, which must agree. For the reversible Metropolis algorithm,
as discussed, this condition is satisfied for any sequence of $i$
(see \eq{e:sequentialMetro}).  If run for a finite number of steps,
the lifted Metropolis algorithm is correct
only if started from a random position $i$.  
It is advantageous to restart this algorithm
after $\lambda\sim N$ time steps by resampling the active sphere $i$.
The chain length $\lambda$ could also be random, sampled
from an appropriate distribution; see the Supplemental Material 4 
for details.
We then observe mixing on a time scale $\bigO{N^2\log N}$, much faster than
all previous Markov chains (see \figmix{}). The $\bigO{N^2}$ time scale
is again brought out by a scaling argument for the discrete Fourier mode
$\Umean$: One chain (sequence of $\sim N$ moves between restarts) of the
lifted Metropolis algorithm can change $\Umean$ by \bigO{1/N^{1/2}}. This
change is of random sign. A random walk in $t/ N$ then yields $\taumix \gtrsim
N^2$.  Simulations indeed show
that restarts every $\lambda \sim N$ time steps yield fastest mixing (see
\fig{figRestartFrequencies}a) and outpace restarts with other scalings with $N$
(see \fig{figRestartFrequencies}b, and below).

The lifted Metropolis algorithm also has a discrete counterpart in the
\emph{lifted TASEP} (see \fig{figAutomatonRules}b): A single sphere (occupied
lattice site) is active and attempts to advance in forward direction. The
sphere remains active if its move is accepted.  Otherwise,  the lifting index
advances to the right-hand neighbor site. The lifted TASEP satisfies global
balance, but without restarts fails to be irreducible.
With restarts every \bigO{N} steps, the lifted
TASEP also mixes on an \bigO{N^2 \log N} time scale.  The infinitesimal limit
of the lifted Metropolis algorithm, $\eps\to0$, is the event-chain algorithm
(ECMC), which also mixes in \bigO{N^2 \log N} lifting moves (see \figmix). It
thus also belongs to the lifted TASEP universality class.

\begin{table}[b]
\begin{tabular}{llc}
    \multicolumn{2}{c}{Local 1D Hard-sphere Markov chains } &  Mixing \\ 
\cline{1-2}
    Continuous & Discrete  & time scale \\ \hline
    Heatbath \cite{Abbrev:RandallWinklerCircle2005}, Metropolis &
    Symm.~SEP \cite{Lacoin_2017_SSEP} & $ N^3 \log N$ 
\\
    Forward \& Lifted Metrop.\footnote{without restarts} & 
    TASEP \cite{BaikLiu2016} & $ 
    N^{5/2}$
\\
    Event-chain, Lifted Metrop. & Lifted TASEP & $ N^2 \log 
    N $
\end{tabular}
\caption{Mixing time scales for local 1D hard-sphere algorithms on the
continuum and on the lattice. The Markov chains on the lowest row all
incorporate restarts. See the Supplemental Material 4 for
pseudo-code implementations. }
\label{t:tabData}
\end{table}

The dynamical universality classes for local 1D hard-sphere algorithms are 
summarized in \tab{t:tabData}. In the lifted TASEP universality class, each 
sphere only moves \bigO{N \log N} times to reach equilibrium, almost saturating 
the lower bound, as \bigO{N} are required to detach each sphere from the compact 
initial state.

The irreversible Markov chains presented here are best viewed from
their deterministic roots, both on the lattice and in the continuum.
Indeed, the lifted TASEP without restarts is a deterministic lattice-gas
automaton satisfying global balance, rather than a Markov chain (see
\fig{figAutomatonRules}b).  Irreducibility and aperiodicity call for an element
of randomness that can be supplied by restarts.
Here, $\lambda\sim N$ (lifted
TASEP) and $\lambda\equiv 1$ (TASEP) represent different universality classes.
In the continuum, the deterministic root is an algorithm without restarts
and invariant step $\eps$, which satisfies global balance.  All presented
Metropolis-type Markov chains may be obtained from this root by resampling
of $i$ or $\eps$.  Algorithms belong to different universality classes,
depending on the resampling rate: Resampling $\eps$ at every step leads to
the lifted Metropolis algorithm without restarts which is in the TASEP class.
The additional resampling of $i$ every $\lambda\sim N$ time steps yields
the lifted Metropolis algorithm with restarts, which is in the lifted 
TASEP class (see \fig{figRestartFrequencies}a, 
the oscillations suggest that resamplings should ideally 
correspond to $\sim N/2$ successfully moved spheres and that multiples of $N$ 
should be avoided). More infrequent restarts ($\lambda \sim N^2$) or more 
frequent ones ($\lambda \sim N^{1/4}$) lead back to the non-lifted TASEP 
class (see \fig{figRestartFrequencies}b). Numerical computations of the 
asymptotic mixing time scales for $\lambda \sim N^\alpha$ with 
$\alpha$ only slightly different from $1$ will require very large 
system sizes in order to overcome the oscillations at $\alpha=1$.

Injecting randomness with a rate $\sim 1/N$ is thus optimal for mixing.  
A particular limit of the continuum root algorithm is ECMC
without restarts.  In 1D, it agrees with Newtonian dynamics with a
single sphere of non-zero velocity, and never mixes.  Resampling $i$ with rate
$\sim 1/N$ (corresponding to an occasional restart of Newtonian dynamics) turns
the deterministic dynamics into a very fast local algorithm, the ECMC, which is
in the lifted TASEP class, and mixes in $\bigO{N^2 \log N}$ lifting moves
(see \figmix).

The infinitesimal displacements of ECMC are crucial for its generalization
to situations where particles simultaneously interact with several
others, such as in more than 1D, or due to long-range forces.
This generalization calls for the factorized Metropolis algorithm
\cite{Michel2014JCP,Peters_2012,KapferKrauth2016}.  The finite-step lifted
Metropolis algorithm remains correct for repulsive interactions restricted to
nearest neighbors.  Indeed, soft repulsive particles with  a $1/x^{12}$
potential reproduce the mixing behavior of hard spheres, with
$\bigO{N^2\log N}$ mixing for the lifted Metropolis and $\bigO{N^{5/2}}$
mixing for lifted Metropolis without restarts (see the Supplemental Material
5 for details).  ECMC has been applied successfully (see
Ref.~\cite{KuKColloquium} for a review), but prior to the present work, its
mixing behavior was not characterized in detail, beyond some partial evidence
for faster mixing time scales \cite{Nishikawa2015}.

In the future, it will be important to clarify how the different universality
classes identified in the present work carry over to higher dimensions and
to what degree they depend on the asymmetry of the step distribution.  Exact
solutions of some of the models in \tab{t:tabData} may be possible. This would
help establish the conceptual framework of lifting and of irreversible Markov
chains beyond the single-particle level \cite{Diaconis2000}.
More generally, the mixing dynamics of hard spheres
from an initial compact state may be interpreted as the equilibration process
of a physical system in response to a sudden change in its Hamiltonian at
time $t=0$.  The analysis of the asymmetric evolution of shock fronts may shed
further light on the new universality class, as previously for the well-studied
TASEP class.  An example for a system strongly out of equilibrium, it will
be interesting to study how entropy production differs between the three
universality classes, and how they may be integrated within the nonequilibrium
fluctuation theorems \cite{EvansSearlesFluctuationTheorem,SeifertSTReview}.

\section{Acknowledgement}

SCK thanks the Institut Philippe Meyer for support, and T.\ Franosch for
helpful discussions. We thank the referees for advice on an earlier version of
the article.

\bibliography{../General.bib,local.bib}

\begin{thebibliography}{32}%
\makeatletter
\providecommand \@ifxundefined [1]{%
 \@ifx{#1\undefined}
}%
\providecommand \@ifnum [1]{%
 \ifnum #1\expandafter \@firstoftwo
 \else \expandafter \@secondoftwo
 \fi
}%
\providecommand \@ifx [1]{%
 \ifx #1\expandafter \@firstoftwo
 \else \expandafter \@secondoftwo
 \fi
}%
\providecommand \natexlab [1]{#1}%
\providecommand \enquote  [1]{``#1''}%
\providecommand \bibnamefont  [1]{#1}%
\providecommand \bibfnamefont [1]{#1}%
\providecommand \citenamefont [1]{#1}%
\providecommand \href@noop [0]{\@secondoftwo}%
\providecommand \href [0]{\begingroup \@sanitize@url \@href}%
\providecommand \@href[1]{\@@startlink{#1}\@@href}%
\providecommand \@@href[1]{\endgroup#1\@@endlink}%
\providecommand \@sanitize@url [0]{\catcode `\\12\catcode `\$12\catcode
  `\&12\catcode `\#12\catcode `\^12\catcode `\_12\catcode `\%12\relax}%
\providecommand \@@startlink[1]{}%
\providecommand \@@endlink[0]{}%
\providecommand \url  [0]{\begingroup\@sanitize@url \@url }%
\providecommand \@url [1]{\endgroup\@href {#1}{\urlprefix }}%
\providecommand \urlprefix  [0]{URL }%
\providecommand \Eprint [0]{\href }%
\providecommand \doibase [0]{http://dx.doi.org/}%
\providecommand \selectlanguage [0]{\@gobble}%
\providecommand \bibinfo  [0]{\@secondoftwo}%
\providecommand \bibfield  [0]{\@secondoftwo}%
\providecommand \translation [1]{[#1]}%
\providecommand \BibitemOpen [0]{}%
\providecommand \bibitemStop [0]{}%
\providecommand \bibitemNoStop [0]{.\EOS\space}%
\providecommand \EOS [0]{\spacefactor3000\relax}%
\providecommand \BibitemShut  [1]{\csname bibitem#1\endcsname}%
\let\auto@bib@innerbib\@empty
\bibitem [{\citenamefont {{Kirkwood}}\ and\ \citenamefont
  {{Monroe}}(1941)}]{KirkwoodMonroe1941}%
  \BibitemOpen
  \bibfield  {author} {\bibinfo {author} {\bibfnamefont {J.~G.}\ \bibnamefont
  {{Kirkwood}}}\ and\ \bibinfo {author} {\bibfnamefont {E.}~\bibnamefont
  {{Monroe}}},\ }\href {\doibase 10.1063/1.1750949} {\bibfield  {journal}
  {\bibinfo  {journal} {\jcp}\ }\textbf {\bibinfo {volume} {9}},\ \bibinfo
  {pages} {514} (\bibinfo {year} {1941})}\BibitemShut {NoStop}%
\bibitem [{\citenamefont {{Hoover}}\ and\ \citenamefont
  {{Ree}}(1968)}]{HooverRee1968}%
  \BibitemOpen
  \bibfield  {author} {\bibinfo {author} {\bibfnamefont {W.~G.}\ \bibnamefont
  {{Hoover}}}\ and\ \bibinfo {author} {\bibfnamefont {F.~H.}\ \bibnamefont
  {{Ree}}},\ }\href {\doibase 10.1063/1.1670641} {\bibfield  {journal}
  {\bibinfo  {journal} {J. Chem. Phys.}\ }\textbf {\bibinfo {volume} {49}},\
  \bibinfo {pages} {3609} (\bibinfo {year} {1968})}\BibitemShut {NoStop}%
\bibitem [{\citenamefont {{Alder}}\ and\ \citenamefont
  {{Wainwright}}(1962)}]{Alder1962}%
  \BibitemOpen
  \bibfield  {author} {\bibinfo {author} {\bibfnamefont {B.~J.}\ \bibnamefont
  {{Alder}}}\ and\ \bibinfo {author} {\bibfnamefont {T.~E.}\ \bibnamefont
  {{Wainwright}}},\ }\href {\doibase 10.1103/PhysRev.127.359} {\bibfield
  {journal} {\bibinfo  {journal} {Phys. Rev.}\ }\textbf {\bibinfo {volume}
  {127}},\ \bibinfo {pages} {359} (\bibinfo {year} {1962})}\BibitemShut
  {NoStop}%
\bibitem [{\citenamefont {Bernard}\ and\ \citenamefont
  {Krauth}(2011)}]{Bernard2011}%
  \BibitemOpen
  \bibfield  {author} {\bibinfo {author} {\bibfnamefont {E.~P.}\ \bibnamefont
  {Bernard}}\ and\ \bibinfo {author} {\bibfnamefont {W.}~\bibnamefont
  {Krauth}},\ }\href {\doibase 10.1103/PhysRevLett.107.155704} {\bibfield
  {journal} {\bibinfo  {journal} {Phys. Rev. Lett.}\ }\textbf {\bibinfo
  {volume} {107}},\ \bibinfo {pages} {155704} (\bibinfo {year}
  {2011})}\BibitemShut {NoStop}%
\bibitem [{\citenamefont {{Asakura}}\ and\ \citenamefont
  {{Oosawa}}(1954)}]{Asakura1954}%
  \BibitemOpen
  \bibfield  {author} {\bibinfo {author} {\bibfnamefont {S.}~\bibnamefont
  {{Asakura}}}\ and\ \bibinfo {author} {\bibfnamefont {F.}~\bibnamefont
  {{Oosawa}}},\ }\href {\doibase 10.1063/1.1740347} {\bibfield  {journal}
  {\bibinfo  {journal} {J. Chem. Phys.}\ }\textbf {\bibinfo {volume} {22}},\
  \bibinfo {pages} {1255} (\bibinfo {year} {1954})}\BibitemShut {NoStop}%
\bibitem [{\citenamefont {Krauth}(2006)}]{SMAC}%
  \BibitemOpen
  \bibfield  {author} {\bibinfo {author} {\bibfnamefont {W.}~\bibnamefont
  {Krauth}},\ }\href {www.smac.lps.ens.fr} {\emph {\bibinfo {title}
  {{Statistical Mechanics: Algorithms and Computations}}}}\ (\bibinfo
  {publisher} {Oxford University Press},\ \bibinfo {year} {2006})\BibitemShut
  {NoStop}%
\bibitem [{\citenamefont {Halperin}\ and\ \citenamefont
  {Nelson}(1978)}]{HalperinNelson1978}%
  \BibitemOpen
  \bibfield  {author} {\bibinfo {author} {\bibfnamefont {B.~I.}\ \bibnamefont
  {Halperin}}\ and\ \bibinfo {author} {\bibfnamefont {D.~R.}\ \bibnamefont
  {Nelson}},\ }\href {\doibase 10.1103/PhysRevLett.41.121} {\bibfield
  {journal} {\bibinfo  {journal} {Phys. Rev. Lett.}\ }\textbf {\bibinfo
  {volume} {41}},\ \bibinfo {pages} {121} (\bibinfo {year} {1978})}\BibitemShut
  {NoStop}%
\bibitem [{\citenamefont {Young}(1979)}]{Young1979VectorCoulomb}%
  \BibitemOpen
  \bibfield  {author} {\bibinfo {author} {\bibfnamefont {A.~P.}\ \bibnamefont
  {Young}},\ }\href {\doibase 10.1103/PhysRevB.19.1855} {\bibfield  {journal}
  {\bibinfo  {journal} {Phys. Rev. B}\ }\textbf {\bibinfo {volume} {19}},\
  \bibinfo {pages} {1855} (\bibinfo {year} {1979})}\BibitemShut {NoStop}%
\bibitem [{\citenamefont {{Alder}}\ and\ \citenamefont
  {{Wainwright}}(1957)}]{Alder1957}%
  \BibitemOpen
  \bibfield  {author} {\bibinfo {author} {\bibfnamefont {B.~J.}\ \bibnamefont
  {{Alder}}}\ and\ \bibinfo {author} {\bibfnamefont {T.~E.}\ \bibnamefont
  {{Wainwright}}},\ }\href {\doibase 10.1063/1.1743957} {\bibfield  {journal}
  {\bibinfo  {journal} {J. Chem. Phys.}\ }\textbf {\bibinfo {volume} {27}},\
  \bibinfo {pages} {1208} (\bibinfo {year} {1957})}\BibitemShut {NoStop}%
\bibitem [{\citenamefont {Alder}\ and\ \citenamefont
  {Wainwright}(1970)}]{AlderWainwrightAutocorrelation1970}%
  \BibitemOpen
  \bibfield  {author} {\bibinfo {author} {\bibfnamefont {B.~J.}\ \bibnamefont
  {Alder}}\ and\ \bibinfo {author} {\bibfnamefont {T.~E.}\ \bibnamefont
  {Wainwright}},\ }\href {\doibase 10.1103/PhysRevA.1.18} {\bibfield  {journal}
  {\bibinfo  {journal} {Phys. Rev. A}\ }\textbf {\bibinfo {volume} {1}},\
  \bibinfo {pages} {18} (\bibinfo {year} {1970})}\BibitemShut {NoStop}%
\bibitem [{\citenamefont {Parisi}\ and\ \citenamefont
  {Zamponi}(2010)}]{Zamponi2010}%
  \BibitemOpen
  \bibfield  {author} {\bibinfo {author} {\bibfnamefont {G.}~\bibnamefont
  {Parisi}}\ and\ \bibinfo {author} {\bibfnamefont {F.}~\bibnamefont
  {Zamponi}},\ }\href {\doibase 10.1103/RevModPhys.82.789} {\bibfield
  {journal} {\bibinfo  {journal} {Rev. Mod. Phys.}\ }\textbf {\bibinfo {volume}
  {82}},\ \bibinfo {pages} {789} (\bibinfo {year} {2010})}\BibitemShut
  {NoStop}%
\bibitem [{\citenamefont {Torquato}\ and\ \citenamefont
  {Stillinger}(2010)}]{TorquatoReview2010}%
  \BibitemOpen
  \bibfield  {author} {\bibinfo {author} {\bibfnamefont {S.}~\bibnamefont
  {Torquato}}\ and\ \bibinfo {author} {\bibfnamefont {F.~H.}\ \bibnamefont
  {Stillinger}},\ }\href {\doibase 10.1103/RevModPhys.82.2633} {\bibfield
  {journal} {\bibinfo  {journal} {Rev. Mod. Phys.}\ }\textbf {\bibinfo {volume}
  {82}},\ \bibinfo {pages} {2633} (\bibinfo {year} {2010})}\BibitemShut
  {NoStop}%
\bibitem [{\citenamefont {{Metropolis}}\ \emph {et~al.}(1953)\citenamefont
  {{Metropolis}}, \citenamefont {{Rosenbluth}}, \citenamefont {{Rosenbluth}},
  \citenamefont {{Teller}},\ and\ \citenamefont {{Teller}}}]{Metropolis1953}%
  \BibitemOpen
  \bibfield  {author} {\bibinfo {author} {\bibfnamefont {N.}~\bibnamefont
  {{Metropolis}}}, \bibinfo {author} {\bibfnamefont {A.~W.}\ \bibnamefont
  {{Rosenbluth}}}, \bibinfo {author} {\bibfnamefont {M.~N.}\ \bibnamefont
  {{Rosenbluth}}}, \bibinfo {author} {\bibfnamefont {A.~H.}\ \bibnamefont
  {{Teller}}}, \ and\ \bibinfo {author} {\bibfnamefont {E.}~\bibnamefont
  {{Teller}}},\ }\href {\doibase 10.1063/1.1699114} {\bibfield  {journal}
  {\bibinfo  {journal} {J. Chem. Phys.}\ }\textbf {\bibinfo {volume} {21}},\
  \bibinfo {pages} {1087} (\bibinfo {year} {1953})}\BibitemShut {NoStop}%
\bibitem [{\citenamefont {Kannan}\ \emph {et~al.}(2003)\citenamefont {Kannan},
  \citenamefont {Mahoney},\ and\ \citenamefont
  {Montenegro}}]{Kannanrapidmixing2003}%
  \BibitemOpen
  \bibfield  {author} {\bibinfo {author} {\bibfnamefont {R.}~\bibnamefont
  {Kannan}}, \bibinfo {author} {\bibfnamefont {M.~W.}\ \bibnamefont {Mahoney}},
  \ and\ \bibinfo {author} {\bibfnamefont {R.}~\bibnamefont {Montenegro}},\
  }in\ \href {\doibase 10.1007/978-3-540-24587-2_68} {\emph {\bibinfo
  {booktitle} {Proc. 14th annual ISAAC}}},\ \bibinfo {series and number}
  {Lecture Notes in Computer Science}\ (\bibinfo  {publisher} {Springer,
  Berlin, Heidelberg},\ \bibinfo {year} {2003})\ pp.\ \bibinfo {pages}
  {663--675}\BibitemShut {NoStop}%
\bibitem [{\citenamefont {Tonks}(1936)}]{Tonks1936}%
  \BibitemOpen
  \bibfield  {author} {\bibinfo {author} {\bibfnamefont {L.}~\bibnamefont
  {Tonks}},\ }\href {\doibase 10.1103/PhysRev.50.955} {\bibfield  {journal}
  {\bibinfo  {journal} {Phys. Rev.}\ }\textbf {\bibinfo {volume} {50}},\
  \bibinfo {pages} {955} (\bibinfo {year} {1936})}\BibitemShut {NoStop}%
\bibitem [{\citenamefont {Randall}\ and\ \citenamefont
  {Winkler}(2005)}]{Abbrev:RandallWinklerCircle2005}%
  \BibitemOpen
  \bibfield  {author} {\bibinfo {author} {\bibfnamefont {D.}~\bibnamefont
  {Randall}}\ and\ \bibinfo {author} {\bibfnamefont {P.}~\bibnamefont
  {Winkler}},\ }\enquote {\bibinfo {title} {Mixing points on a circle},}\ in\
  \href@noop {} {\emph {\bibinfo {booktitle} {Approximation, Randomization and
  Combinatorial Optimization}}},\ \bibinfo {series} {Lecture Notes in Computer
  Science}, Vol.\ \bibinfo {volume} {3624},\ \bibinfo {editor} {edited by\
  \bibinfo {editor} {\bibnamefont {{C. Chekuri et al.}}}}\ (\bibinfo
  {publisher} {Springer, Berlin, Heidelberg},\ \bibinfo {year}
  {2005})\BibitemShut {NoStop}%
\bibitem [{\citenamefont {Kapfer}\ and\ \citenamefont
  {Krauth}(2017{\natexlab{a}})}]{SUPP}%
  \BibitemOpen
  \bibfield  {author} {\bibinfo {author} {\bibfnamefont {S.~C.}\ \bibnamefont
  {Kapfer}}\ and\ \bibinfo {author} {\bibfnamefont {W.}~\bibnamefont
  {Krauth}},\ }\href@noop {} {} (\bibinfo {year} {2017}{\natexlab{a}}),\
  \bibinfo {note} {supplemental Material XXXX}\BibitemShut {NoStop}%
\bibitem [{\citenamefont {Gwa}\ and\ \citenamefont
  {Spohn}(1992)}]{GwaSpohnPRL1992}%
  \BibitemOpen
  \bibfield  {author} {\bibinfo {author} {\bibfnamefont {L.-H.}\ \bibnamefont
  {Gwa}}\ and\ \bibinfo {author} {\bibfnamefont {H.}~\bibnamefont {Spohn}},\
  }\href {\doibase 10.1103/PhysRevLett.68.725} {\bibfield  {journal} {\bibinfo
  {journal} {Phys. Rev. Lett.}\ }\textbf {\bibinfo {volume} {68}},\ \bibinfo
  {pages} {725} (\bibinfo {year} {1992})}\BibitemShut {NoStop}%
\bibitem [{\citenamefont {Chou}\ \emph {et~al.}(2011)\citenamefont {Chou},
  \citenamefont {Mallick},\ and\ \citenamefont {Zia}}]{ChouTASEP2011}%
  \BibitemOpen
  \bibfield  {author} {\bibinfo {author} {\bibfnamefont {T.}~\bibnamefont
  {Chou}}, \bibinfo {author} {\bibfnamefont {K.}~\bibnamefont {Mallick}}, \
  and\ \bibinfo {author} {\bibfnamefont {R.~K.~P.}\ \bibnamefont {Zia}},\
  }\href {http://stacks.iop.org/0034-4885/74/i=11/a=116601} {\bibfield
  {journal} {\bibinfo  {journal} {Rep. Prog. Phys.}\ }\textbf {\bibinfo
  {volume} {74}},\ \bibinfo {pages} {116601} (\bibinfo {year}
  {2011})}\BibitemShut {NoStop}%
\bibitem [{\citenamefont {Baik}\ and\ \citenamefont {Liu}(2016)}]{BaikLiu2016}%
  \BibitemOpen
  \bibfield  {author} {\bibinfo {author} {\bibfnamefont {J.}~\bibnamefont
  {Baik}}\ and\ \bibinfo {author} {\bibfnamefont {Z.}~\bibnamefont {Liu}},\
  }\href {\doibase 10.1007/s10955-016-1665-y} {\bibfield  {journal} {\bibinfo
  {journal} {J. Stat. Phys.}\ }\textbf {\bibinfo {volume} {165}},\ \bibinfo
  {pages} {1051} (\bibinfo {year} {2016})}\BibitemShut {NoStop}%
\bibitem [{\citenamefont {Bernard}\ \emph {et~al.}(2009)\citenamefont
  {Bernard}, \citenamefont {Krauth},\ and\ \citenamefont
  {Wilson}}]{Bernard2009}%
  \BibitemOpen
  \bibfield  {author} {\bibinfo {author} {\bibfnamefont {E.~P.}\ \bibnamefont
  {Bernard}}, \bibinfo {author} {\bibfnamefont {W.}~\bibnamefont {Krauth}}, \
  and\ \bibinfo {author} {\bibfnamefont {D.~B.}\ \bibnamefont {Wilson}},\
  }\href {\doibase 10.1103/PhysRevE.80.056704} {\bibfield  {journal} {\bibinfo
  {journal} {Phys. Rev. E}\ }\textbf {\bibinfo {volume} {80}},\ \bibinfo
  {pages} {056704} (\bibinfo {year} {2009})}\BibitemShut {NoStop}%
\bibitem [{\citenamefont {Chen}\ \emph {et~al.}(1999)\citenamefont {Chen},
  \citenamefont {Lovász},\ and\ \citenamefont {Pak}}]{Abbrev:Chen1999}%
  \BibitemOpen
  \bibfield  {author} {\bibinfo {author} {\bibfnamefont {F.}~\bibnamefont
  {Chen}}, \bibinfo {author} {\bibfnamefont {L.}~\bibnamefont {Lovász}}, \
  and\ \bibinfo {author} {\bibfnamefont {I.}~\bibnamefont {Pak}},\ }\href@noop
  {} {\bibfield  {journal} {\bibinfo  {journal} {Proc. 17th Ann. ACM Symp.
  Theory of Computing}\ ,\ \bibinfo {pages} {275}} (\bibinfo {year}
  {1999})}\BibitemShut {NoStop}%
\bibitem [{\citenamefont {Diaconis}\ \emph {et~al.}(2000)\citenamefont
  {Diaconis}, \citenamefont {Holmes},\ and\ \citenamefont
  {Neal}}]{Diaconis2000}%
  \BibitemOpen
  \bibfield  {author} {\bibinfo {author} {\bibfnamefont {P.}~\bibnamefont
  {Diaconis}}, \bibinfo {author} {\bibfnamefont {S.}~\bibnamefont {Holmes}}, \
  and\ \bibinfo {author} {\bibfnamefont {R.~M.}\ \bibnamefont {Neal}},\
  }\href@noop {} {\bibfield  {journal} {\bibinfo  {journal} {Ann. Appl.
  Probab.}\ }\textbf {\bibinfo {volume} {10}},\ \bibinfo {pages} {726}
  (\bibinfo {year} {2000})}\BibitemShut {NoStop}%
\bibitem [{\citenamefont {{Michel}}\ \emph {et~al.}(2014)\citenamefont
  {{Michel}}, \citenamefont {{Kapfer}},\ and\ \citenamefont
  {{Krauth}}}]{Michel2014JCP}%
  \BibitemOpen
  \bibfield  {author} {\bibinfo {author} {\bibfnamefont {M.}~\bibnamefont
  {{Michel}}}, \bibinfo {author} {\bibfnamefont {S.~C.}\ \bibnamefont
  {{Kapfer}}}, \ and\ \bibinfo {author} {\bibfnamefont {W.}~\bibnamefont
  {{Krauth}}},\ }\href {\doibase 10.1063/1.4863991} {\bibfield  {journal}
  {\bibinfo  {journal} {J. Chem. Phys.}\ }\textbf {\bibinfo {volume} {140}},\
  \bibinfo {eid} {054116} (\bibinfo {year} {2014})}\BibitemShut {NoStop}%
\bibitem [{\citenamefont {Kapfer}\ and\ \citenamefont
  {Krauth}(2017{\natexlab{b}})}]{KuKColloquium}%
  \BibitemOpen
  \bibfield  {author} {\bibinfo {author} {\bibfnamefont {S.~C.}\ \bibnamefont
  {Kapfer}}\ and\ \bibinfo {author} {\bibfnamefont {W.}~\bibnamefont
  {Krauth}},\ }\href@noop {} {} (\bibinfo {year} {2017}{\natexlab{b}}),\
  \bibinfo {note} {to appear in EPJB}\BibitemShut {NoStop}%
\bibitem [{\citenamefont {{Nishikawa}}\ \emph {et~al.}(2015)\citenamefont
  {{Nishikawa}}, \citenamefont {{Michel}}, \citenamefont {{Krauth}},\ and\
  \citenamefont {{Hukushima}}}]{Nishikawa2015}%
  \BibitemOpen
  \bibfield  {author} {\bibinfo {author} {\bibfnamefont {Y.}~\bibnamefont
  {{Nishikawa}}}, \bibinfo {author} {\bibfnamefont {M.}~\bibnamefont
  {{Michel}}}, \bibinfo {author} {\bibfnamefont {W.}~\bibnamefont {{Krauth}}},
  \ and\ \bibinfo {author} {\bibfnamefont {K.}~\bibnamefont {{Hukushima}}},\
  }\href {\doibase 10.1103/PhysRevE.92.063306} {\bibfield  {journal} {\bibinfo
  {journal} {Phys. Rev. E}\ }\textbf {\bibinfo {volume} {92}},\ \bibinfo {eid}
  {063306} (\bibinfo {year} {2015})}\BibitemShut {NoStop}%
\bibitem [{\citenamefont {Aldous}\ and\ \citenamefont
  {Diaconis}(1986)}]{AldousDiaconis1986}%
  \BibitemOpen
  \bibfield  {author} {\bibinfo {author} {\bibfnamefont {D.}~\bibnamefont
  {Aldous}}\ and\ \bibinfo {author} {\bibfnamefont {P.}~\bibnamefont
  {Diaconis}},\ }\href {http://www.jstor.org/stable/2323590} {\bibfield
  {journal} {\bibinfo  {journal} {Am. Math. Monthly}\ }\textbf {\bibinfo
  {volume} {93}},\ \bibinfo {pages} {333} (\bibinfo {year} {1986})}\BibitemShut
  {NoStop}%
\bibitem [{\citenamefont {Lacoin}(2017)}]{Lacoin_2017_SSEP}%
  \BibitemOpen
  \bibfield  {author} {\bibinfo {author} {\bibfnamefont {H.}~\bibnamefont
  {Lacoin}},\ }\href {\doibase 10.1214/16-aihp759} {\bibfield  {journal}
  {\bibinfo  {journal} {Ann I H Poincar{\'{e}}-Pr}\ }\textbf {\bibinfo {volume}
  {53}},\ \bibinfo {pages} {1402} (\bibinfo {year} {2017})}\BibitemShut
  {NoStop}%
\bibitem [{\citenamefont {Peters}\ and\ \citenamefont
  {de~With}(2012)}]{Peters_2012}%
  \BibitemOpen
  \bibfield  {author} {\bibinfo {author} {\bibfnamefont {E.~A. J.~F.}\
  \bibnamefont {Peters}}\ and\ \bibinfo {author} {\bibfnamefont
  {G.}~\bibnamefont {de~With}},\ }\href {\doibase 10.1103/PhysRevE.85.026703}
  {\bibfield  {journal} {\bibinfo  {journal} {Phys. Rev. E}\ }\textbf {\bibinfo
  {volume} {85}},\ \bibinfo {pages} {026703} (\bibinfo {year}
  {2012})}\BibitemShut {NoStop}%
\bibitem [{\citenamefont {Kapfer}\ and\ \citenamefont
  {Krauth}(2016)}]{KapferKrauth2016}%
  \BibitemOpen
  \bibfield  {author} {\bibinfo {author} {\bibfnamefont {S.~C.}\ \bibnamefont
  {Kapfer}}\ and\ \bibinfo {author} {\bibfnamefont {W.}~\bibnamefont
  {Krauth}},\ }\href {\doibase 10.1103/PhysRevE.94.031302} {\bibfield
  {journal} {\bibinfo  {journal} {Phys. Rev. E}\ }\textbf {\bibinfo {volume}
  {94}},\ \bibinfo {pages} {031302} (\bibinfo {year} {2016})}\BibitemShut
  {NoStop}%
\bibitem [{\citenamefont {Evans}\ and\ \citenamefont
  {Searles}(2002)}]{EvansSearlesFluctuationTheorem}%
  \BibitemOpen
  \bibfield  {author} {\bibinfo {author} {\bibfnamefont {D.~J.}\ \bibnamefont
  {Evans}}\ and\ \bibinfo {author} {\bibfnamefont {D.~J.}\ \bibnamefont
  {Searles}},\ }\href {\doibase 10.1080/00018730210155133} {\bibfield
  {journal} {\bibinfo  {journal} {Adv. Phys.}\ }\textbf {\bibinfo {volume}
  {51}},\ \bibinfo {pages} {1529} (\bibinfo {year} {2002})}\BibitemShut
  {NoStop}%
\bibitem [{\citenamefont {Seifert}(2012)}]{SeifertSTReview}%
  \BibitemOpen
  \bibfield  {author} {\bibinfo {author} {\bibfnamefont {U.}~\bibnamefont
  {Seifert}},\ }\href {\doibase 10.1088/0034-4885/75/12/126001} {\bibfield
  {journal} {\bibinfo  {journal} {Rep. Prog. Phys.}\ }\textbf {\bibinfo
  {volume} {75}},\ \bibinfo {pages} {126001} (\bibinfo {year}
  {2012})}\BibitemShut {NoStop}%
\end{thebibliography}%
\ifincludesupplements

\clearpage

\title{Supplemental material to ``Irreversible local Markov chains with rapid
convergence towards equilibrium''}

\maketitle

\onecolumngrid
\renewcommand{\theequation}{S\arabic{equation}}
\renewcommand{\thefigure}{S\arabic{figure}}
\setcounter{equation}{0}
\setcounter{figure}{0}
\setcounter{page}{1}
\newcommand\varu{\ensuremath{\operatorname{Var}u}}

\subsection{Supplemental Item 1: Balance conditions, irreducibility, 
aperiodicity}

The mixing of a Markov chain describes how a \quot{worst-case} initial
probability distribution $\pi^{t=0}$ converges towards the equilibrium
distribution $\pi^{t = \infty} = \pi$.  More generally, the time evolution of
the distribution $\pi^t$ is given by
\begin{equation} 
\pi^t = T \cdot \pi^{t-1},
\end{equation}
where $T$ is the transfer matrix composed of the algorithmic transition
probabilities $\ptrans(a\to b)$. This can be expressed as:
\begin{equation} 
    \pi^t(b) = \sum_a \pi^{t-1}(a) \ptrans(a \to b).
    \label{EQU_GlobalBalance_t_tp_Probs}
\end{equation} 
The steady state, for $ t \to \infty$, is defined by $\pi^t = \pi^{t-1} = \pi$,
and for the hard-sphere system considered in this work it equals $1$ for each
valid configuration.  For \eq{EQU_GlobalBalance_t_tp_Probs} to be valid, the
global balance condition must hold:
\begin{equation}
    \pi(b) = \sum_a \pi(a) \ptrans(a \to b) = \sum_a \FCAL_{a \to b}.
\label{EQU_GlobalBalanceInFlow} 
\end{equation} 
The global-balance condition thus states that the probability flow from all
configurations $a$ (including $b$) into $b$ must equal $\pi(b)$. Because
of $\sum_c \ptrans(b \to c) =1$ (if the Markov chain continues, it must go
somewhere starting from $b$, maybe even back to $c= b$), it can also be written
as
\begin{equation}
\sum_a \FCAL_{a \to b} = \pi(b) = \sum_c \FCAL_{b \to c},
\label{EQU_GlobalFlowInFlowOut} 
\end{equation} 
that is, the flow into $b$ equals the flow out of $b$. 
The detailed-balance condition
\begin{equation}
    \pi(b) \ptrans(b \to a) = \pi(a) \ptrans(a \to b)
    \equiv
    \FCAL_{b \to a}  = \FCAL_{a \to b}
    \quad \forall a,b
\end{equation}
implies global balance, but it is more restrictive on the algorithmic
transition probabilities $\ptrans(a \to b)$.  It implies that the steady-state
flow is reversible and that the net flow $\FCAL_{a \to b} - \FCAL_{b \to a}$
between any $a$ and $b$ vanishes.

For the lifted Metropolis algorithm treated in this work, the configurations
are defined by the lifting variables in addition to the \quot{physical}
configurations. The global balance condition then requires that the flow into
a \quot{lifted} configuration $(a,i)$ equals the weight of $(a,i)$, which is
$\pi[(a,i)] = \pi(a) = 1$.

Besides the global balance condition, the Markov chain must satisfy two further
conditions in order to be sure to converge to the probability distribution
$\pi$ in the limit of infinite times. The \emph{irreducibility condition}
requires that for every pair of states $a, b$, there exists a natural number $m$
such that $(T^m)_{ab} > 0$, i.\,e., if the Markov chain starts at $a$,
with $\pi^0(a)=1$,
there is a finite probability $\pi^m(b)$.  This condition (in the physics
literature often called \quot{ergodicity}) ensures that there is only a single
stationary solution of \eq{EQU_GlobalBalanceInFlow}, and that it is reached
from an arbitrary initial condition. The event-chain algorithm without restarts
violates the irreducibility condition, because after each event, the position
$x_i$ of sphere $i$ comes to equal $x_{i+1} - d$. For $d=0$, for example,
the set of the $N-1$ positions at event times does not change with time.
Nevertheless, restarts reinstall the irreducibility. Finally, the Markov chain
must satisfy the \emph{aperiodicity condition}, which means that it should be
free of cycles. All Markov chains discussed in this work are aperiodic.

\subsection{Supplemental Item 2: Rigorous and operational definitions of mixing 
time}

Rigorously, a Markov chain's approach to equilibrium is quantified through
the total variation distance between the finite-time probability distributions
$\pi^t$ and the stationary solution $\pi = \pi^{t=\infty}$:
\begin{equation}
    || \pi^t - \pi || = \frac{1}{2} \sum_j | \pi^t(j) - \pi(j)|.
\label{EQU_TotalVariation}
\end{equation}
Here, $\pi^t = T^t \pi^0$ depends on the initial distribution $\pi^ 0$ chosen.
The total variation distance of \eq{EQU_TotalVariation} approaches zero as
$t \to \infty$, which is equivalent to the statement that the Markov chain
defined by $T$ mixes. The mixing time, for a given tolerance $\eps$, is defined
as the time at which the total variation distance drops below $\eps$ from a
\quot{worst-case} initial state $\pi^{\text{w.c.}}$:
\begin{equation}
    \max_i || T^t \pi^{\text{w.c.}}  - \pi || \leq \eps \quad \text{for all
$t\geq\taumix(\eps)$}.
\end{equation}
The mixing time thus determines the time it takes to obtain a first (single)
sample of the equilibrium distribution.  The mixing time is usually larger
than the correlation time (given by the inverse spectral gap of the transfer
matrix), the time it takes to move from one equilibrium configuration to
another.

For the Markov chains discussed in this work, we do not compute the mixing time
rigorously and rather suppose that the \quot{worst-case} initial distribution
is realized when all the spheres $i, i+1, \dots, N,N+1, \dots, N+i-1$,
are compactly spaced by the sphere diameter $d$.  We then track the approach
towards equilibrium through what we suppose to be the slowest variable, namely
the half-system distance $u$ between any particle $i$ and particle $i + N/2$
(with periodic boundary conditions applied):
\begin{equation}
    u = \underbrace{\delta_{i+N/2} +\dotsc+ \delta_{i+1}}_\text{$N/2$ terms}.
\end{equation}
Because of periodic boundary conditions, thermal averages of this quantity
do not depend on $i$.  The probability distribution $\pi(u)$ can be computed
exactly (see \cite{SMAC}):
\begin{align}
    \pi(u)
&=
    \frac{\Gamma(N-1)}{\Lfree^{N-1}} 
    \frac{(\Lfree u-u^2)^{N/2-1}}{(N/2-1)!^2}.
\end{align}
We obtain
\begin{align}
    \langle u \rangle 
&=
    \Lfree / 2, 
\notag    
\\
    \langle u^2 \rangle
&=
    \frac{\Lfree^2 (N+2)}{4(N+1)},
\notag    
\\
    \varu
&=
    \frac{N^2\lfree^2}{4 (N+1)}.
\label{eqVaruExactMean}
\end{align}
In the compact initial configuration, \varu\ takes the value
\begin{equation*}
    \operatorname{Var}_\text{compact}(u)
        = \langle u^2 \rangle_\text{compact} - \langle u 
\rangle^2_\text{compact}
        = \frac {N^2 \lfree^2 }{4}.
\label{eqVaruExactCompact}
\end{equation*}
In \fig{figMixing}c of the main text, we track the ratio of the empirical
variance of the half-system distance and of the exact equilibrium variance
given by \eq{eqVaruExactMean}. We note that this ratio has to go from \bigO{N}
at $t=0$ towards $\sim 1$ at $t \sim \taumix$.

\subsection{Supplemental Item 3: Density independence of Heatbath and Metropolis 
algorithms}

In higher dimensions, a Monte Carlo algorithm for hard-sphere systems
is defined for moves $\xvec_i \to \xvec_i + \epsvec$ that must satisfy a
non-overlap condition which forces the centers of spheres $i$ and $j$ (both of
diameter $d$) to remain distant by more than $d$. In one dimension, one may
complement this by an ordering constraint $x_i < x_{i+1}$ (with appropriate
periodic boundary conditions). A move can thus be rejected for two reasons:
\begin{enumerate}
 \item A physical overlap of two spheres, 
 \item A violation of the imposed ordering of spheres (an inversion). 
\end{enumerate}
These two conditions are respected if a move $x_i \to x_i + \eps$ (with 
$\eps$ larger or smaller than $0$)  is accepted only if 
\begin{equation}
\underbrace{x_{i-1} - x_{i} + d}_{\text{$-\delta_i$, 
\quot{left} gap variable}} <  \eps < 
\underbrace{x_{i+1} - x_{i} - d}_{\text{$\delta_{i+1}$, 
\quot{right} gap variable}}.
\end{equation}
It follows that two starting configurations with different values of the
$x_i$ and $d$ but the same gap variables $\delta_i$ perform equivalently
under the same Monte Carlo dynamics. Furthermore, two starting configurations
and Monte Carlo algorithms with gap variables and distribution $\proba(\eps)$
scaled by the same factor also realize the same dynamics. It also follows from
this  simple analysis of the Monte Carlo dynamics that the one-dimensional
hard-sphere model cannot have a phase transition (as the sphere diameter $d$
does not modify the dynamics) and that the dynamics, at constant $N$, is
independent of the  density  $N d / L$ if the distribution of $\eps$ scales
with $\lfree$ (for example, if $\eps$ is a uniform random number in $[0; 2.5
\lfree]$, as used in the main text).

\subsection{Supplemental Item 4: Pseudo-code implementations of the
1D hard-sphere Metropolis algorithms}

In the following pseudo-code implementations of the algorithms discussed
in the main text, we suppose real-valued ordered positions $x_i < x_{i+1}$.
We identify particles $i$ and $i\pm N$, and implicitly apply periodic boundary
conditions for the positions, $x_i + L \equiv x_{i+N}$.

\subsubsection{Reversible Metropolis algorithm}
 
The reversible Metropolis algorithm satisfies detailed balance, as discussed
in the main text. The algorithm remains correct for arbitrary interaction
potentials if the move from $a$ towards a trial configuration $b$ is accepted
with the standard Metropolis acceptance filter $\min[1, \pi(b) / \pi(a)]$
(\cite{Metropolis1953}, see  \cite{SMAC}).
\begin{align}
\parbox{.8\linewidth}{
    \texttt{Initialize $a=\{x_1, \dotsc, x_N\}$}\\
    \texttt{For $t=1, \dotsc, t_{\max}$:}\\
    \hspace*{2em}\texttt{Sample $i$ uniformly from $\{1, \dotsc, N\}$}\\
    \hspace*{2em}\texttt{Sample $\eps>0$ from $\proba(\eps)$}\\
    \hspace*{2em}\texttt{Sample $\sigma$ uniformly from $\{\pm 1\}$}\\
    \hspace*{2em}\texttt{If $x_{i-1} + d < x_i + \sigma\eps < x_{i+1} - d$:} 
(Accept move)\\
    \hspace*{2em}\hspace*{2em}\texttt{Set $x_i := x_i + \sigma\eps$}\\
    \hspace*{2em}\texttt{Measure observables} ($a$ is distributed according to 
$\pi$ for $t \to \infty$)\\
}
    \label{eqReversibleMetroAlgo}
\end{align}

\subsubsection{Sequential Metropolis algorithm}

The sequential Metropolis algorithm performs a sequence of one-particle
updates which individually satisfy detailed balance, but the sequence as a
whole violates detailed balance. As discussed in the main text, the sequential
Metropolis, introduced in \cite{Metropolis1953}, is the historically first
irreversible Markov chain. The increasing sequence of spheres used in the
following pseudo-code implementation can be replaced any other sequence
that updates each sphere $i$ with finite frequency. The sequential Metropolis
algorithm remains correct for arbitrary interaction potentials if the move from
$a$ towards a trial configuration $b$ is accepted with the standard Metropolis
acceptance filter $\min[1, \pi(b) / \pi(a)]$ (\cite{Metropolis1953}, see
\cite{SMAC}).
\begin{align}
\parbox{.8\linewidth}{
    \texttt{Initialize $a=\{x_1, \dotsc, x_N\}$}\\
    \texttt{For $t=1, \dotsc, t_{\max}$:}\\
    \hspace*{2em}\texttt{Set $i := t \operatorname{mod} N$} \\
    \hspace*{2em}\texttt{Sample $\eps>0$ from $\proba(\eps)$}\\
    \hspace*{2em}\texttt{Sample $\sigma$ uniformly from $\{\pm 1\}$}\\
    \hspace*{2em}\texttt{If $x_{i-1} + d < x_i + \sigma\eps < x_{i+1} - d$:} 
(Accept move)\\
    \hspace*{2em}\hspace*{2em}\texttt{Set $x_i := x_i + \sigma\eps$}\\
    \hspace*{2em}\texttt{Measure observables} ($a$ is distributed according to 
$\pi$ for $t \to \infty$)\\
}
\end{align}

\subsubsection{Forward Metropolis algorithm}

The forward Metropolis algorithm moves spheres only in positive direction
($\sigma =+1$). As discussed in the main text, the uniform random sampling
of the sphere index $i$ is essential to insure that this algorithm satisfies
the global-balance condition (for example, we prove in the main text that the
sequential forward Metropolis algorithm is incorrect.)
\begin{align}
\parbox{.8\linewidth}{
    \texttt{Initialize $a=\{x_1, \dotsc, x_N\}$}\\
    \texttt{For $t=1, \dotsc, t_{\max}$:}\\
    \hspace*{2em}\texttt{Sample $i$ uniformly from $\{1, \dotsc, N\}$}\\
    \hspace*{2em}\texttt{Sample $\eps>0$ from $\proba(\eps)$}\\
    \hspace*{2em}\texttt{If $ x_i + \eps < x_{i+1} - d$:} 
(Accept move)\\
    \hspace*{2em}\hspace*{2em}\texttt{Set $x_i := x_i + \eps$}\\
    \hspace*{2em}\texttt{Measure observables} ($a$ is distributed according to 
$\pi$ for $t \to \infty$)\\
}
    \label{eqForwardAlgo}
\end{align}

\subsubsection{Lifted Metropolis algorithm (without restarts)}

In the lifted Metropolis algorithm without restarts, all spheres move in
positive direction $\sigma = +1$. Only the active sphere $i$ can move, and $i$
remains active after a move of $i$ that was accepted.  Otherwise, the lifting
move $i \to i+1$ ensues. Initially, $i$ must be chosen uniformly among all
spheres, although this is only relevant for small running times $t_{\max}$.
As illustrated in the following pseudo-code implementation, both the physical
move and the lifting move count as one time step. The algorithm is irreducible
for a general distribution of steps $\proba(\eps)$.
\begin{align}
\parbox{.8\linewidth}{
    \texttt{Initialize $a=\{x_1, \dotsc, x_N\}$}\\
    \texttt{Sample $i$ uniformly from $\{1, \dotsc, N\}$}\\
    \texttt{For $t=1, \dotsc, t_{\max}$:}\\
    \hspace*{2em}\texttt{Sample $\eps>0$ from $\proba(\eps)$}\\
    \hspace*{2em}\texttt{If $ x_i + \eps < x_{i+1} - d$:} 
(Physical move)\\
    \hspace*{2em}\hspace*{2em}\texttt{Set $x_i := x_i + \eps$}\\
    \hspace*{2em}\texttt{Else:} (Lifting move)\\
    \hspace*{2em}\hspace*{2em}\texttt{Set $i := (i + 1) \operatorname{mod} 
N$}\\
    \hspace*{2em}\texttt{Measure observables} ($a$ is distributed according to 
$\pi$ for $t \to \infty$)\\
}
\end{align}

\subsubsection{Lifted Metropolis algorithm (with restarts)}

The lifted Metropolis algorithm with restarts is obtained from the previous
lifted Metropolis algorithm by resampling the active sphere $i$ (uniformly
among all spheres) after $\lambda$ time steps.  (The below algorithm
continues to satisfy global balance if the same $\eps$ is chosen for the
whole chain $c$, rather than to resample it at each time step).  $\lambda$
may be a random positive integer, or a constant.  For best efficiency, the
distribution of chain lengths $\proba(\lambda)$ should be chosen such that
$\langle\lambda\rangle_\proba \sim N$, i.\,e., a typical chain traverses an
extensive part of the system.  This ensures that the algorithm is in the lifted
TASEP class.
\begin{align}
\parbox{.8\linewidth}{
    \texttt{Initialize $a=\{x_1, \dotsc, x_N\}$}\\
    \texttt{For $c=1, \dotsc, C$:}\\
    \hspace*{2em}\texttt{Sample $\lambda$ from $\proba(\lambda)$}\\
    \hspace*{2em}\texttt{Sample $i$ uniformly from $\{1, \dotsc, N\}$}\\
    \hspace*{2em}\texttt{For $m=1, \dotsc, \lambda$:}\\
    \hspace*{2em}\hspace*{2em}\texttt{Sample $\eps>0$ from $\proba(\eps)$}\\
    \hspace*{2em}\hspace*{2em}\texttt{If $ x_i + \eps < x_{i+1} - d$:} 
(Physical move)\\
    \hspace*{2em}\hspace*{2em}\hspace*{2em}\texttt{Set $x_i := x_i + \eps$}\\
    \hspace*{2em}\hspace*{2em}\texttt{Else:} (Lifting move)\\
    \hspace*{2em}\hspace*{2em}\hspace*{2em}\texttt{Set $i := (i + 1) 
                           \operatorname{mod} N$}\\
    \hspace*{2em}\hspace*{2em}\texttt{Measure observables} ($a$ is distributed 
according to $\pi$ for $t \to \infty$)\\
}
    \label{eqLiftedAlgo}
\end{align}

\subsubsection{Event-chain algorithm}

The hard-sphere \emph{event-chain Monte Carlo algorithm} \cite{Bernard2009}
remains valid in arbitrary dimensions. It generalizes to arbitrary interactions
using the factorized Metropolis acceptance rule \cite{Michel2014JCP}. For
simplicity, the following pseudo-code only treats one-dimensional hard
spheres, where it presents the continuum limit $\eps\to 0$, $\lambda
\to \infty$, $\eps \lambda = \ell$ of the lifted Metropolis with restarts
\eq{eqLiftedAlgo}. Restarts are essential for irreducibility of the algorithm.
The distribution of chain lengths $\ell$ should be chosen such that
$\langle\ell\rangle_\proba \sim \Lfree$, i.\,e., a typical chain traverses an
extensive part of the system.  This ensures that the algorithm is in the lifted
TASEP class.
\begin{align}
\parbox{.8\linewidth}{
    \texttt{Initialize $a=\{x_1, \dotsc, x_N\}$}\\
    \texttt{For $c=1, \dotsc, C$:}\\
    \hspace*{2em}\texttt{Sample $\ell>0$ from $\proba(\ell)$}\\
    \hspace*{2em}\texttt{Sample $i$ uniformly from $\{1, \dotsc, N\}$}\\
    \hspace*{2em}\texttt{While $\ell > 0$:}\\
    \hspace*{2em}\hspace*{2em}\texttt{Set $\Delta := \min(x_{i+1} -x_i - d, \ell)$}\\
    \hspace*{2em}\hspace*{2em}\texttt{Set $x_i := x_i + \Delta$}\\
    \hspace*{2em}\hspace*{2em}\texttt{Set $\ell := \ell - \Delta$}\\
    \hspace*{2em}\hspace*{2em}\texttt{Set $i := (i + 1) \operatorname{mod} N$}\\
    \hspace*{2em}\texttt{Measure observables} ($a$ is distributed according to 
$\pi$ for $t \to \infty$)\\
}
    \label{eqEcAlgo}
\end{align}

\subsection{Supplemental Item 5: Lifted forward Metropolis algorithm
with soft potentials}

The global-balance algorithms discussed in the main text (with finite $\eps$)
remain valid under the condition that the interaction is restricted to only
nearest neighbors, is monotonous and that it enforces the ordering constraint
$x_{i+1} > x_i$. Under these conditions, the weight $\pi(a)$ can be written as
\begin{equation}
 \pi(a) = \prod_{m=1}^{N} \pitilde(x_{m+1} - x_m),
\end{equation}
where periodic boundary conditions are again assumed, and the function
$\pitilde(\Delta x)$ monotonically increases with $\Delta x$ and is zero
for $\Delta x < 0$. In addition, the move from a configuration $a$ (with
positions $x$) to a configuration $b$ (with positions $x'$) is accepted with
the factorized Metropolis filter \cite{Michel2014JCP}
\begin{equation}
 \pacc(a \to b) =  \prod_m  \min \glc 1, 
 \frac{\pitilde(x'_{m+1} - x'_m)}
{\pitilde(x_{m+1} - x_m)} \grc.
\label{e:factorized_weight}
\end{equation}
For the considered move $x_i \to x_i' = x_i + \eps$ with $\eps > 0$, only two
terms in \eq{e:factorized_weight} have to be considered (and only one of them
contributes), namely
\begin{equation}
\min \glc 1, 
 \frac{\pitilde(x_{i} + \eps - x_{i-1})}
{\pitilde(x_{i} - x_{i-1})} \grc = 1
\quad \text{and} \quad
\min \glc 1, 
 \frac{\pitilde(x_{i+1} - x_i -\eps)}
{\pitilde(x_{i+1} - x_i)} \grc =
 \frac{\pitilde(x_{i+1} - x_i -\eps)}
{\pitilde(x_{i+1} - x_i)} 
\leq 1.
\label{e:factorized_weight2}
\end{equation}
Note that because of the monotonicity of the weights,
$\pitilde(x_{i} + \eps - x_{i-1}) \geq \pitilde(x_{i} - x_{i-1})$, etc.
It follows that the move $x_i \to x_i + \eps$ is accepted with probability
\begin{equation}
\pacc(x_i \to x_i + \eps) = \frac{\pitilde(x_{i+1} - x_i -\eps)} {\pitilde(x_{i+1} - x_i)}.
\end{equation}
As in the main text, we have to show that the flow 
$\FCAL_{(a,i)}^{\text{lift}}$
into a lifted configuration
$(a,i)$ (either due to an accepted move of sphere $i$ or to a 
rejected move of sphere $i-1$) satisfies
\begin{equation}
\FCAL_{(a,i)}^{\text{lift}} = \ACAL^+_i + \RCAL^+_{i-1} = \pi(a), 
    \label{eqFinalrSoftBalance}
\end{equation}
where $\ACAL^+_i = \int\mathrm d \eps \proba(\eps) \ACAL^+_i(\eps)$, and
analogous for $\RCAL^+_{i-1}$.
But this follows because, for any fixed $\eps > 0$, the accepted flow
corresponds to a move of sphere $i$ from $x_i - \eps$ to $x_i$:
\begin{align*}
\ACAL^+_i(\eps) &= \pi( \dots,x_{i-1}, x_i - \eps, x_{i+1}, \dots) \pacc(x_i - \eps \to 
x_i)\\
          &= \glc \dots \pitilde(x_i - x_{i-1} - \eps)
          \pitilde(x_{i+1} - x_i + \eps)
          \pitilde(x_{i+2} - x_{i+1}) \dots \grc \frac{\pitilde(x_{i+1} - x_i)}
{\pitilde(x_{i+1} - x_i + \eps)}
\end{align*}
and because the rejected flow corresponds to a (rejected) move of sphere 
$i-1$ from $x_{i-1}$ to $x_{i-1} + \eps$:
\begin{align*}
\RCAL^+_{i-1}(\eps) &= \pi( \dots,x_{i-1}, x_i, x_{i+1}, \dots) \glc 1 -  \pacc(x_{i-1} 
\to   x_{i-1} + \eps) \grc  \\
          &= \glc \dots \pitilde(x_i - x_{i-1})
          \pitilde(x_{i+1} - x_i) \pitilde(x_{i+2} - x_{i+1}) \dots \grc 
          \glc 1 - \frac{\pitilde(x_i - x_{i-1} - \eps)}
          {\pitilde(x_i - x_{i-1})} \grc
\end{align*}
Thus \eq{eqFinalrSoftBalance} holds irrespective of the distribution
$\proba(\eps)$.

We have implemented soft-potential version of the Metropolis algorithms
discussed in the main text for a $1/x^{12}$ potential, and confirmed
that without restarts, it belongs to the TASEP universality class, and
with restarts every $\lambda \sim N$ steps, to the lifted TASEP class, see
Fig.~\ref{figSoftMixing} panels b) and c).  In the same system, the reversible
Metropolis algorithm mixes as $\bigO{N^3\log N}$, just as in hard spheres.  The
severe conditions on the applicability of the lifted Metropolis finite-$\eps$
algorithm disappear in the limit $\eps \to 0$, that is, for the generalized
event-chain algorithm \cite{Michel2014JCP}, see Fig.~\ref{figSoftMixing}a).

\begin{figure}[htb]
\includegraphics[width=\linewidth]{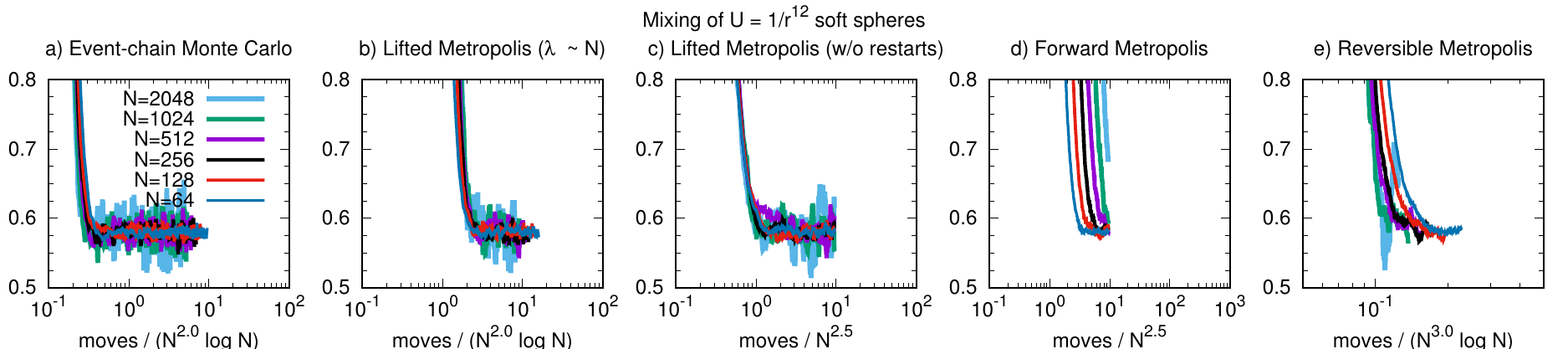}
\caption{Mixing of Metropolis-type Markov chains with a $1/x^{12}$ soft
potential.  We plot the quantity $\operatorname{Var} u$ on the vertical
axis, and rescale the number of moves on the horizontal axis to extract the
scaling of the mixing time $\taumix$ with $N$.  a) and b) With restarts every
$\lambda\sim N$ moves, the lifted Metropolis and event-chain algorithms are in
the lifted TASEP class.  c) The lifted Metropolis without restarts is in the
TASEP class, $\taumix = \bigO{N^{5/2}}$.  d) The forward Metropolis algorithm
for soft spheres mixes slower than \bigO{N^{5/2}}, apparently requiring
logarithmic corrections.  e) The reversible Metropolis algorithm with a
factorized filter mixes in $\bigO{N^3\log N}$.  For this system, $\tilde\pi(x)
= \exp(-\Gamma x^{-12})$ for $x>0$ and $\tilde\pi(x) = 0$ otherwise.
}
\label{figSoftMixing}
\end{figure}
\fi
\end{document}